\documentclass[reprint,twocolumn,superscriptaddress,nofootinbib,amsmath,amssymb]{aastex63}
\usepackage{graphicx}% Include figure files
\usepackage{dcolumn}% Align table columns on decimal point
\usepackage{hyperref}% add hypertext capabilities
\hypersetup{
  colorlinks=true,        % false: boxed links; true: colored links
  linkcolor=blue,         % color of internal links
  citecolor=blue,         % color of links to bibliography
  urlcolor=blue           % color of external links
}
\usepackage{amsmath}
\usepackage{times}

\newcommand{\ie}{i.e.,~}
\newcommand{\eg}{e.g.,~}

\begin{document}

\title[A scale-independent sound speed in neutron stars]{A general,
  scale-independent description of the sound speed in neutron stars}

\author[0000-0002-8669-4300]{Christian Ecker}
\affiliation{Institut f\"ur Theoretische Physik, Goethe Universit\"at,
  Max-von-Laue-Str. 1, 60438 Frankfurt am Main, Germany}

\author[0000-0002-1330-7103]{Luciano Rezzolla}

\affiliation{Institut f\"ur Theoretische Physik, Goethe Universit\"at,
  Max-von-Laue-Str. 1, 60438 Frankfurt am Main, Germany}
\affiliation{School of Mathematics, Trinity College, Dublin 2, Ireland}
\affiliation{Frankfurt Institute for Advanced Studies,
  Ruth-Moufang-Str. 1, 60438 Frankfurt am Main, Germany}

%\date{\today}
 
\begin{abstract}
\noindent
Using more than a million randomly generated equations of state that
satisfy theoretical and observational constraints we construct a novel,
scale-independent description of the sound speed in neutron stars where
the latter is expressed in a unit-cube spanning the normalised radius,
$r/R$, and the mass normalized to the maximum one, $M/M_{\rm TOV}$. From
this generic representation, a number of interesting and surprising
results can be deduced. In particular, we find that light (heavy) stars
have stiff (soft) cores and soft (stiff) outer layers, respectively, or
that the maximum of the sound speed is located at the center of light
stars but moves to the outer layers for stars with $M/M_{\rm
  TOV}\gtrsim0.7$, reaching a constant value of $c_s^2=1/2$ as $M\to
M_{\rm TOV}$. We also show that the sound speed decreases below the
conformal limit $c_s^2=1/3$ at the center of stars with $M=M_{\rm
  TOV}$. Finally, we construct an analytic expression that accurately
describes the radial dependence of the sound speed as a function of the
neutron-star mass, thus providing an estimate of the maximum sound speed
expected in a neutron star.
\end{abstract}

%\maketitle

\section{Introduction}

The extreme conditions in the neutron-star interiors pose a formidable
problem for the theoretic modelling of nuclear matter several times
denser than the saturation density of atomic nuclei $n_s:=0.16\,\rm
fm^{-3}$. While effective field-theory calculations are arguably the most
important tool to obtain theoretical predictions for the behavior of
dense matter, the associated uncertainties become large at densities
several times $n_s$, such as those present in neutron-star cores. In
addition, first-principle perturbative Quantum-Chromodynamics (QCD)
calculations are only reliable at densities much larger than those
realized in the neutron-star interior, but provide important consistency
conditions for the modelling of matter at lower
densities~\citep{Fraga2014, Annala2019, Komoltsev:2021jzg,Annala:2022,
  Gorda:2022jvk, Somasundaram:2022ztm}. Our theoretical control of even
basic quantities such as the equation of state (EOS) of dense nuclear and
quark matter that, in the simplest case, is a relation between pressure
and the energy density $p(e)$, is therefore still very limited, often
forcing the use of agnostic approaches to build the EOS of nuclear matter
at neutron-star densities. At the same time, recent and upcoming
observations of neutron stars and their merger events represent an unique
opportunity to gain information about strongly coupled dense matter under
conditions that are either difficult or impossible to create in
experiments.

It is therefore important to combine our current theoretical knowledge
with the available observational data to make predictions for the EOS and
related quantities that determine the macroscopic properties of neutron
stars. One such quantity is the (adiabatic) sound speed, which is
  defined as the derivative of the pressure with respect to the energy
  density at fixed entropy per baryon $s$~\citep{Rezzolla_book:2013}
\begin{equation}
  \label{eq:soundspeed}
  c_s^2:=\left(\frac{ \partial p}{ \partial e}\right)_s\,.
\end{equation}
Among its many important properties, the sound speed provides a measure
of the stiffness of matter, or in other words, it determines the amount
of material pressure available to balance the gravitational pull and
therefore prevent a neutron star from collapsing to a black
hole. Clearly, a large speed of sound corresponds to a large stiffness,
which, in turn, allows for the support of neutron stars with large radii
$R$ and large maximum masses $M_{\rm TOV}$. A similarly simple logic
would suggest that the sound speed should reach its maximum value in the
core of the star. 

There exist a number of works~\citep{Moustakidis2017, Tews2018a,
  Margaritis2020, Kanakis-Pegios:2020kzp, Hippert:2021, Altiparmak:2022}
addressing the question whether the sound speed as a function of density
in QCD has an upper limit that is smaller than the speed of light. One
natural conjecture for this bound is the value in conformally symmetric
matter $c_s^2=1/3$ \citep[see, \eg][]{Bedaque2015, Alsing2017}), such as
realized in QCD at asymptotically large density. However, it turns out
assuming this conformal limit ($c_s^2\leq 1/3$) strictly at all densities
leads to a strong tension with astrophysical measurements of neutron-star
masses $M\gtrsim2~M_{\odot}$~\citep{Antoniadis2013, Cromartie2019,
  Fonseca2021}, which favor stiff EOSs with $c_s^2\gtrsim 1/3$ at
densities $\gtrsim n_s$. In addition, various theoretic approaches, such
as QCD at large isospin density~\citep{Carignano:2016lxe}, two-color QCD
\citep{Hands:2006ve}, quarkyonic matter \citep{McLerran:2018hbz,
  Margueron:2021dtx, Duarte:2021tsx}, models for high-density QCD~\citep{
  Pal:2021qav, Ma:2021zev, Braun:2022olp, Leonhardt:2019fua} and models based on the
gauge/gravity duality~\citep{Ecker:2017fyh, Demircik:2021zll,
  Kovensky:2021kzl} predict $c_s^2>1/3$ at finite
densities.

In other words, there now exists evidence and consensus that the sound
speed at neutron-star densities exceeds the conformal limit. However, it
still remains unclear \textit{what} should this maximum value be and,
more importantly, \textit{where} in the neutron-star interior it is
achieved. For instance, it is natural to expect that the maximum sound
speed should always take place at the center of the star, as this is
where the largest densities are achieved. As we will show below, this
simple logic is valid for light stars but fails spectacularly as one
considers stars near the maximum mass.

In this \textit{Letter} we investigate the behavior of the sound speed in
the neutron-star interior using input for the EOS from nuclear theory and
perturbative QCD and impose observational constraints on
neutron-star masses, radii and tidal deformabilities. Our goal is to make
statements that are universal in the sense that they do not depend on a
particular choice of the EOS or on any of the macroscopic scales such as
the radius and maximum mass of a given EOS. To achieve such a model
independence, we randomly generate more than a million EOSs that are by
construction consistent with nuclear theory and perturbative QCD at low and high
densities, respectively, and satisfy observational constraints of
isolated and binary neutron-star merger measurements. We then generate
probability density functions (PDF) for the sound speed inside
non-rotating neutron stars and extract their median and $95$\% confidence
intervals. By choosing dimensionless coordinates for the radial ($r/R$)
and mass dependence ($M/M_{\rm TOV}$), we obtain a novel and entirely
scale-independent description of the sound speed in the neutron-star
interior.

\section{Methods}
\label{sec:methods}  

Our setup is similar to the one presented by~\citet{Altiparmak:2022} and
we briefly review it here. The starting point is the construction of a
large number of EOSs, which we achieve by patching together several
different components. At the lowest densities, \ie for $n<0.5\,n_s$, we
use the Baym-Pethick-Sutherland (BPS) model~\citep{Baym71b} to describe
the neutron-star crust. In the range $0.5\,n_s<n<1.1\,n_s$ we randomly
sample polytropes to cover the entire range between the soft and stiff
EOSs of \citet{Hebeler:2013nza}. At large densities, $\gtrsim 40\,n_s$,
corresponding to a fixed baryon chemical potential of $\mu=2.6\,{\rm
  GeV}$, we impose the perturbative QCD result of \citet{Fraga2014} for the pressure
$p(X,\mu)$ of cold quark matter in terms of a renormalization scale
parameter $X$, which we sample uniformly in the range
$[1,4]$. Although such high densities are not realized in neutron
  stars~\citep[see, \eg][]{Altiparmak:2022}, imposing $p(X,\mu)$
  constrains the EOS at neutron-star densities. Finally, in the
intermediate regime of densities, i.e., $1.1\,n_s < n \lesssim 40\,n_s$,
we use the parametrization method introduced by~\citet{Annala2019}, which
uses a continuous combination of piecewise-linear segments for the sound
speed as a function of the chemical potential $c_s^2(\mu)$ as a starting
point to construct thermodynamic quantities
\begin{equation}
  \label{eq:cs2}
  c_s^2(\mu)=\frac{\left(\mu _{i+1}-\mu \right) c_{s,i}^2+\left(\mu -\mu
   _i\right) c_{s,i+1}^2{}}{\mu _{i+1}-\mu _i}\,,
\end{equation}
where $\mu_i$ and $c_{s,i}^2$ are parameters of the $i$-th segment in the
range $\mu_i\le \mu \le \mu_{i+1}$ (throughout this article we use units
in which the speed of light and Newton's constant are equal to one,
$c=G=1$). The number density can then be expressed as
\begin{equation}
  \label{eq:n}
 n(\mu)=n_1\,\exp \left({\int_{\mu_1}^\mu \frac{ d \mu^\prime}{\mu^\prime
     c_s^2(\mu^\prime)}}\right)\,,
\end{equation}
where $n_1=1.1\,n_s$ and $\mu_1=\mu(n_1)$ is fixed by the corresponding
chemical potential of the polytrope. Integrating Eq.~\eqref{eq:n} then
gives the pressure
\begin{equation}\label{eq:p}
 p(\mu)=p_1+\int_{\mu_1}^\mu  d \mu^\prime n(\mu^\prime)\,,
\end{equation}
where the integration constant $p_1$ is fixed by the pressure of the
polytrope at $n=n_1$. We integrate Eq.~\eqref{eq:p} numerically, using a
fixed number of $7$ segments of the form \eqref{eq:cs2} for the sound
speed. Following the procedure above, we construct a large number of EOSs
by choosing randomly the maximum speed of sound $c_{s,{\rm
    max}}^2\in[0,1]$ and uniformly sample the remaining free coefficients
$\mu_i\in[\mu_1,\mu_{N+1}]$, where $\mu_{N+1}=2.6\,{\rm GeV}$, and
$c_{s,i}^2\in[0,c_{s,{\rm max}}^2]$ in their respective domains. In this
way, we construct $8\times 10^6$ EOSs that are consistent with
theoretical uncertainties in nuclear theory and perturbative QCD.

In order to impose constraints provided by the astronomical observations,
we solve for each EOS the Tolman-Oppenheimer-Volkoff (TOV) equations and
keep only those EOSs that are consistent with the mass measurements of
J0348+0432~\citep{Antoniadis2013} ($M=2.01\pm 0.04\,M_{\odot}$) and
of J0740+6620~\citep{Cromartie2019,Fonseca2021} ($M=2.08 \pm
0.07\, M_{\odot}$) by rejecting those with maximum mass $M_{_{\rm
    TOV}}<2.0\,M_{\odot}$. In addition, we impose the radius measurements
by NICER of J0740+6620~\citep{Miller2021, Riley2021} and of
J0030+0451~\citep{Riley2019,MCMiller2019b} by rejecting EOSs with
$R<10.75\,{\rm km}$ at $M=2.0\,M_{\odot}$ and $R<10.8\,{\rm km}$ at
$M=1.1\,M_{\odot}$, respectively. Finally, we exploit the detection of
GW171817 by LIGO/Virgo to set an upper bound on the binary tidal
deformability $\tilde \Lambda<720$ (low-spin priors) \citep{Abbott2018a}.
Denoting respectively with $M_{i}$, $R_{i}$, and $\Lambda_{i}$ the
masses, radii, and tidal deformabilities of the binary components, where
$\Lambda_i=\tfrac{2}{3}k_2\left({R_i}/{M_i}\right)^5$, $i=1,2$, and $k_2$
is the second tidal Love number, we compute the binary tidal
deformability as
\begin{equation}
\tilde{\Lambda} := \frac{16}{13} \frac{ \left( 12M_2 + M_1\right) M_{1}^4
  \Lambda_1 + \left(12M_1 + M_2 \right)M_{2}^4 \Lambda_2}{\left( M_1 +
  M_2\right)^5}\,.
\end{equation}
For any choice of $M_{1,2}$ and $R_{1,2}$, we then reject those EOSs with
$\tilde\Lambda>720$ for a chirp mass $\mathcal{M}_{\rm chirp}:=(M_1
M_2)^{3/5}(M_1+M_2)^{-1/5}=1.186\,M_{\odot}$ and $q:={M_2}/{M_1}>0.73$ as
required for consistency with LIGO/Virgo data for
GW170817~\citep{Abbott2018a}. Imposing the observational constraints
reduces our original set of $8\times 10^6$ EOSs to $\approx 10^6$ viable
EOSs that form the basis for the results presented in the next section.

We note that, in principle, there exist also estimates for the upper
bound on the maximum mass $M_{\rm TOV}\lesssim
2.16^{+0.17}_{-0.15}\,M_{\odot}$~(see \citet{Rezzolla2017} but also
\citet{Margalit2017, Ruiz2017, Shibata2019, Nathanail2021}). However,
since this bound requires a number of uncertain model assumptions about
the kilonova modeling of GRB170817A emitted by the merger event GW170817,
we do not impose them on the results presented in the main text, but
rather study their impact in the Appendix~\ref{sec:appendix_a}.

\section{Results}
\label{sec:results}

\begin{figure}[htb]
\center
\includegraphics[width=0.99\columnwidth]{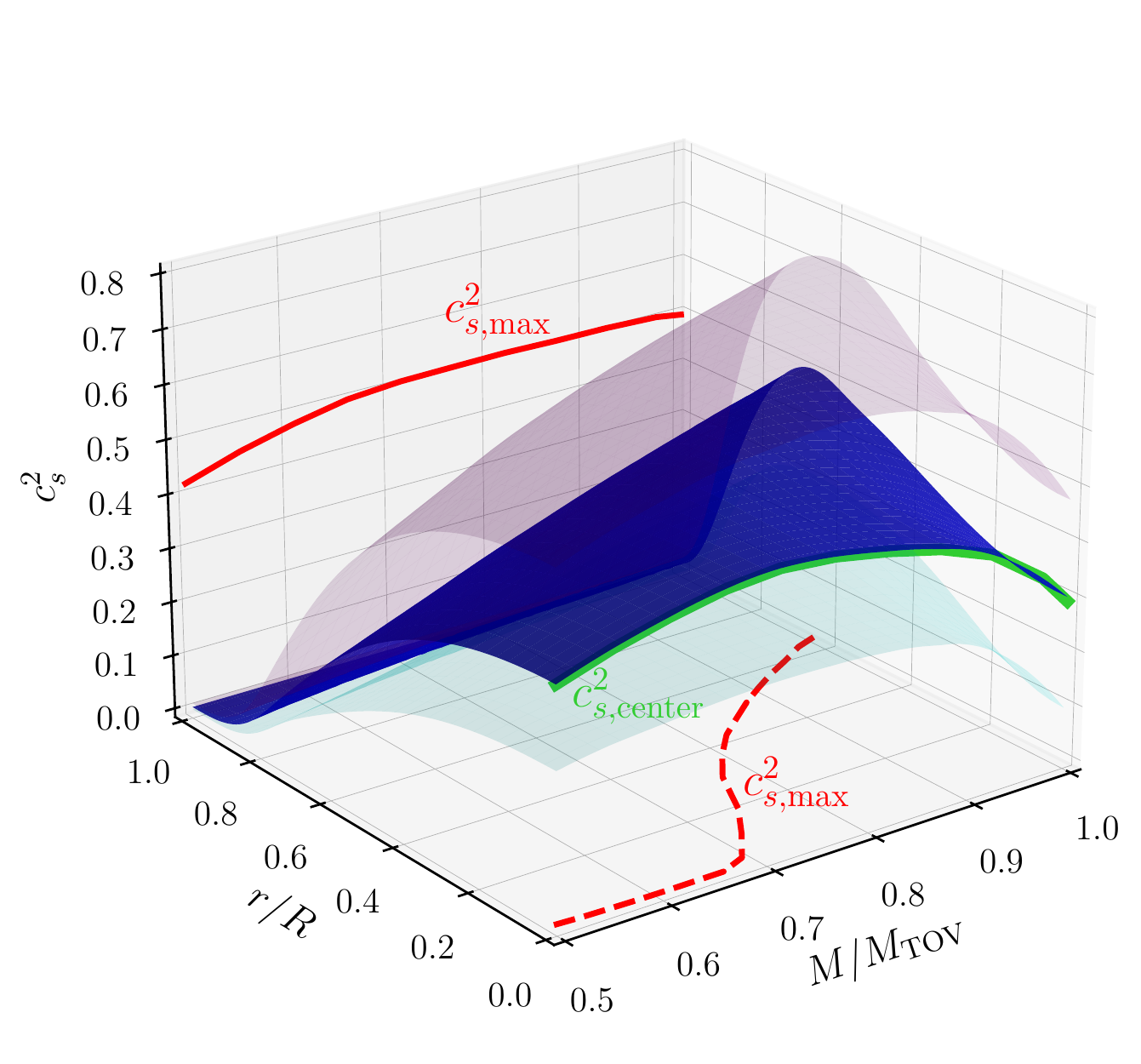}
\caption{Scale-independent description of the sound speed as a function
  of the mass $M$ normalized by the maximum mass $M_{\rm TOV}$ and of the
  radial location $r$ normalized by the neutron-star radius $R$. The blue
  surface represents the median of the PDF, while the cyan and purple
  surfaces the lower and upper $95$\% credibility intervals,
  respectively. Different lines are used also to show important
  properties of the PDF: the red solid line shows the value of the
  maximum sound speed, $c^2_{s,{\rm max}}$, as a function of the stellar
  mass, while the red dashed line shows the location within the star of
  $c^2_{s,{\rm max}}$ as a function of the stellar mass; the green solid
  line reports the sound speed at the stellar center, $c^2_{s,{\rm
      center}}$, as a function of the stellar mass.}
\label{fig:cs2_3D}
\end{figure}

Figure~\ref{fig:cs2_3D} represents the synthesis and the essence of our
novel scale-independent representation of the sound speed in neutron
stars, which is described in a unit cube having as coordinates: the
normalised radius $r/R\in[0,1]$, the normalised mass $M/M_{\rm
  TOV}\in[0,1]$, and the (normalised) sound speed squared $c_s^2\in[0,1]$
(to aid the visualisation of the data we restrict the unit cube to the
most interesting region). In this unit cube, we report with the blue
surface the median of the sound-speed-squared PDF, while the cyan and
purple surfaces mark the $95$\% credibility intervals.

\begin{figure*}
\center
\includegraphics[width=1\textwidth]{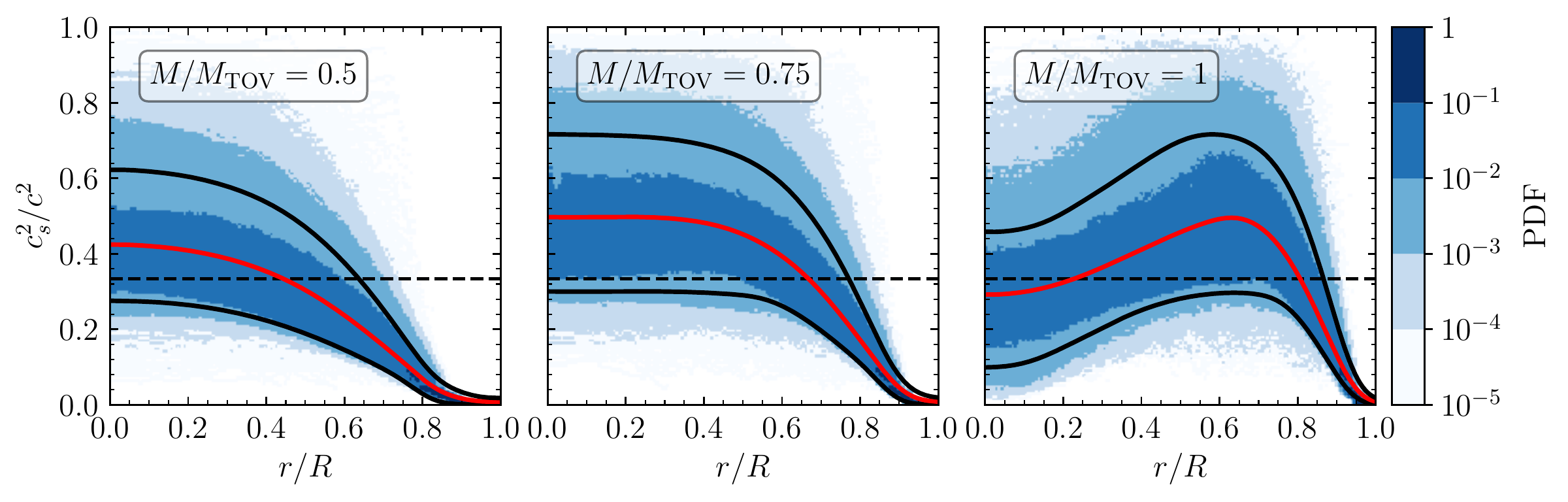}
\includegraphics[width=1\textwidth]{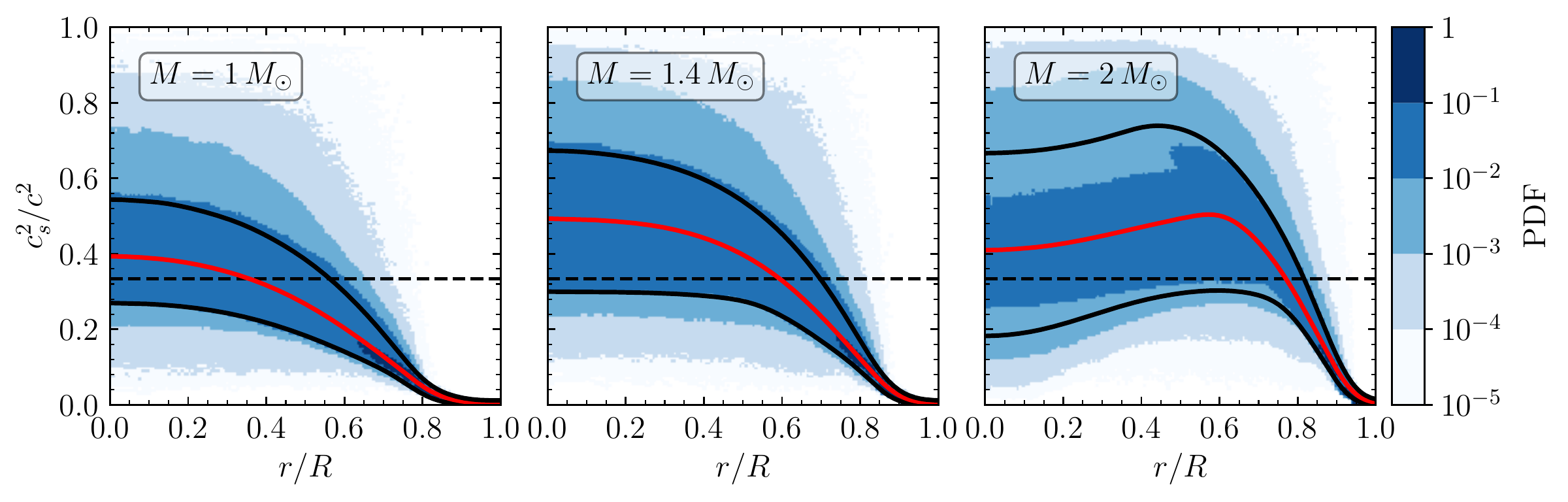}
\caption{\textit{Top Panels: } Probability density functions (PDFs) for
  the sound speed as a function of the normalized radial coordinate $r/R$
  for three fixed values of the masses $M=0.5\,M_{\rm TOV}$ (left),
  $M=0.75\,M_{\rm TOV}$ (middle) and $M=M_{\rm TOV}$ (right). Red lines
  represent the median of the distribution, black solid lines are the
  upper and lower bound of the $95$\% credibility interval, while black
  dashed lines indicate the conformal limit $c_s^2=1/3$. \textit{Bottom
    Panels: } The same as in the top panels but shown for representative
  stellar masses in units of the solar mass: $M=1.0\,M_{\odot}$ (left),
  $M=1.4\,M_{\odot}$ (middle) and $M=2.0\,M_{\odot}$ (right).}
\label{fig:pdfMtov}
\end{figure*}

\begin{figure}
\center
\includegraphics[width=0.95\columnwidth]{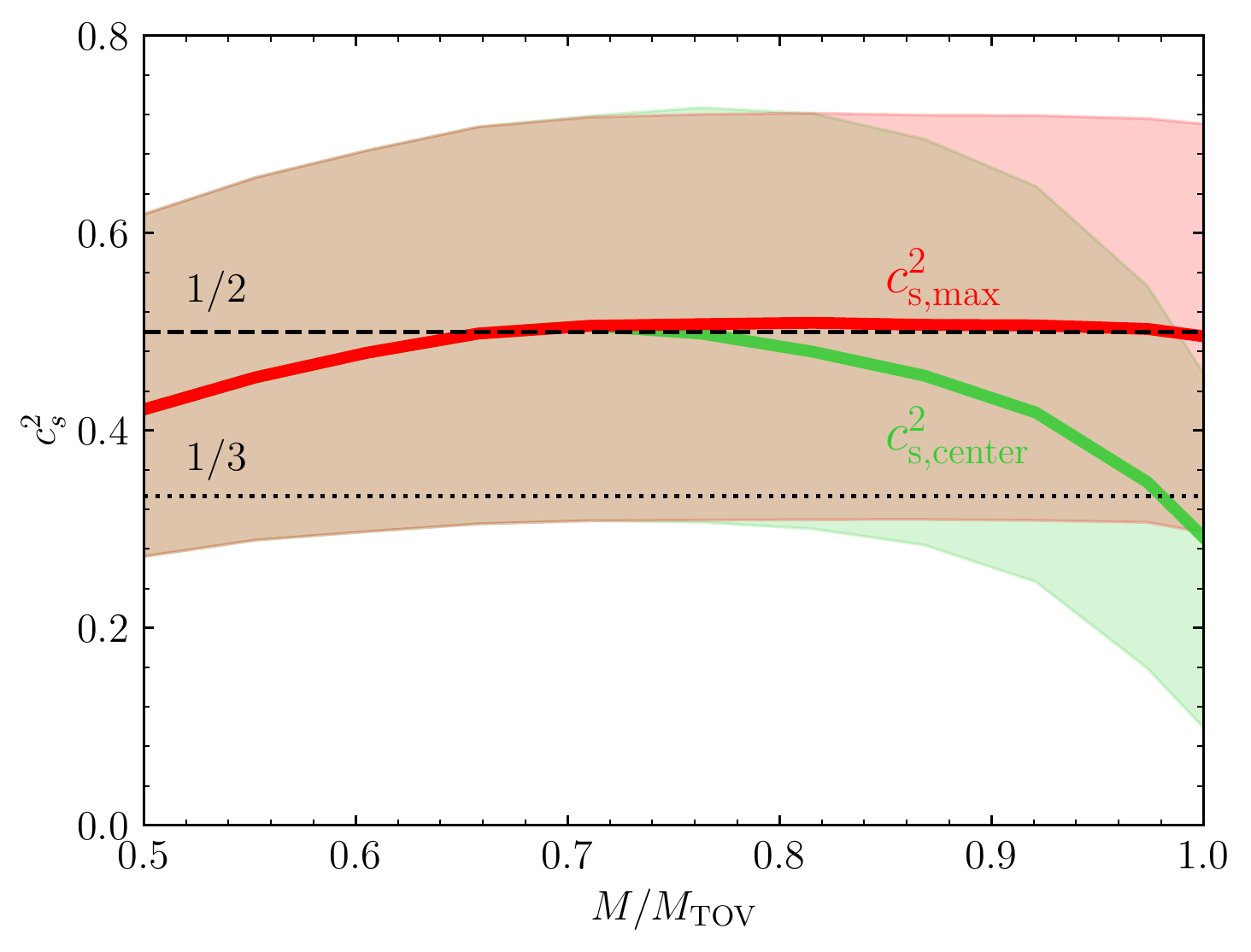}
\caption{Mass dependence of the median of the local maximum sound speed
  $c^2_{s,{\rm max}}$ (red line) and of the sound speed at the
  neutron-star center $c^2_{s,{\rm center}}$ (green line); the red- and
  green-shaded areas indicate the corresponding $95$\% credibility
  intervals. Black dashed and dotted lines mark half the speed of light
  and the conformal limit, respectively. }
\label{fig:cs2maxM}
\end{figure}

We first discuss the most prominent global features of the median sound
speed. Obviously, close to the surface of the stars ($r/R\approx 1$) the
sound speed is small $c^2_s \sim \mathcal{O}(10^{-2})$ and approximately
independent of the mass. This is because the underlying nuclear theory
description at low density is tightly constrained and has small sound
speed. Moving inside the star ($r/R\lesssim 0.8$), the sound speed
develops a non-trivial mass dependence. For $M/M_{\rm TOV}\gtrsim 0.7$,
the sound speed changes from a monotonic to a non-monotonic function of
$r/R$ that has a single local maximum, $c^2_{s, {\rm max}}$, as shown by
the red dashed and solid lines in Fig.~\ref{fig:cs2_3D}. Importantly, the
radial location of the maximum sound speed depends on the mass (red
dashed line), so that it is at the center of the stars ($r/R=0$) for
light stars ($M/M_{\rm TOV} \lesssim 0.7$) but then moves to the outer
layers of the stars ($r/R \simeq 0.5-0.7$) for heavy stars ($M/M_{\rm
  TOV} \gtrsim 0.7$). This interesting and somewhat surprising behaviour
highlights how the structure of a compact star depends sensitively on its
mass. As we will further discuss below, the value of this local maximum
becomes independent of the mass for sufficiently large masses ($M/M_{\rm
  TOV} \gtrsim 0.7$) and attains a constant value of $c_{s,\rm
  max}^2\approx 1/2$ (see also Fig.~\ref{fig:cs2maxM}). Another important
feature of the sound speed is the non-monotonic behavior of the value at
the center of the stars, $c^2_{s,{\rm center}}$, as a function of the
mass, as shown by the green solid line in Fig.~\ref{fig:cs2_3D}. Note
that the maximum value of $c^2_{s,{\rm center}}$ is reached by stars with
intermediate mass ($M/M_{\rm TOV}\approx 0.7$), while the minimum value
is not obtained in the lightest stars ($M/M_{\rm TOV}=0.5$), but rather
in the heaviest stars ($M/M_{\rm TOV}=1$), where it is even below the
conformal limit, \ie $c^2_{s,{\rm center}} \simeq 0.3$ (see also
top-right panel of Fig.~\ref{fig:pdfMtov}).

In Fig.~\ref{fig:pdfMtov} we show the PDFs for three different values of
the dimensionless mass parameter (top panels) and for three values of the
mass when expressed in solar masses (bottom panels). In all cases, the
red lines represent the median of the PDFs, while the black solid lines
the corresponding $95$\% credibility intervals and black dashed lines
mark the conformal limit $c_s^2=1/3$. Note how the red curves clearly
show the transition from a monotonic to a non-monotonic radial dependence
of the sound speed when going from light (left) to heavy stars
(right). The sound speed in light stars increases relatively slow from
the surface towards the center, reaching the conformal value only roughly
at half its outer radius, \ie $r/R\approx 0.5$.

As a result, the outer layers of light stars are relatively soft and are
therefore more sensitive to tidal disruptions than heavier stars. In
stars with intermediate mass ($M/M_{\rm TOV}\approx 0.75$) the sound
speed increases more rapidly towards the center and remains at a constant
value $c_s^2\approx c_{s,\rm max}^2=1/2$ over a large portion
($r/R\lesssim 0.4$) of the core region. Hence, these stars have a
relatively large and stiff core compared to lighter ones. Finally, the
heaviest stars ($M/M_{\rm TOV}\approx 1$) do not have a sound speed that
changes monotonically within the star, but, as mentioned above, develop a
local maximum rather far from the center ($r/R\approx 0.65$). Hence,
heavy stars have relatively soft cores, but very stiff outer layers. The
stiffening in the outer layers clearly is a consequence of imposing the
two solar-mass constraint $M_{\rm TOV}>2\,M_{\odot}$, while the softening
in the core is required to satisfy the perturbative QCD boundary conditions at large
densities and, at the same time, maintain causality at all
densities. A physically intuitive way of looking at this
  behaviour is the following: as the stellar core softens, the burden of
  keeping the heavy star stable against gravitational collapse has to be
  taken by the outer layers, that have therefore to become stiff and with
  large sound speeds. Hence, the appearance of a local maximum in the
  sound speed is a direct consequence of the interplay between the
  astrophysical constraints ($M_{\rm TOV}\gtrsim 2\,M_{\odot}$) and the
  perturbative QCD constraints at high densities.

We note that traditional nuclear-theory EOSs that only capture confined
nuclear matter typically do not take into account the perturbative QCD constraints at
large densities and therefore cannot lead to the soft core that our
statistical approach points out. Indeed, widely used EOSs built on the
present understanding of low-density nuclear theory typically predict
high-mass stars with cores that are systematically stiffer and therefore
with larger sounds speed (see Fig.~\ref{fig:compare} in
Appendix~\ref{sec:appendix_b}).

The panels in Fig.~\ref{fig:pdfMtov} also make it very easy to appreciate
that while light and intermediate-mass stars have super-conformal central
sound speeds ($c^2_{s,{\rm center}} > 1/3$; black dashed lines), heavy
stars have sub-conformal central sound speeds ($c^2_{s,{\rm center}} <
1/3)$, reaching a local minimum of $c_{s,\rm center}^2\approx 0.3$ for
$M/M_{\rm TOV} = 1$.

The smooth behaviour of the medians of the sound speed shown in the
panels of Fig.~\ref{fig:pdfMtov} encourages the representation of this
behaviour in terms of analytic functions, namely
\begin{equation}
  \label{eq:fitcs2}
	c_s^2(x)= \left(\alpha e^{\beta x^2}+\gamma
        e^{\delta\left(x-\epsilon\right)^2}\right)\left[1-{\rm
            tanh}\left(\zeta x-\eta\right)\right]\,,
\end{equation}
where $x:=r/R$ and $\alpha=0.18, \beta=0.14,\gamma=0.71, \delta=-6.30,
\epsilon=-1.5, \zeta=10.00, \eta=8.9$ are the fitting parameters for the
most interesting case, namely the median of the PDF when $M/M_{\rm
  TOV}=1$. For completeness, Table \ref{tab:fit} in
Appendix~\ref{sec:appendix_c} reports the values of these coefficients
for different mass cuts and different credibility intervals.

\begin{figure}
  \center
  \includegraphics[width=0.95\columnwidth]{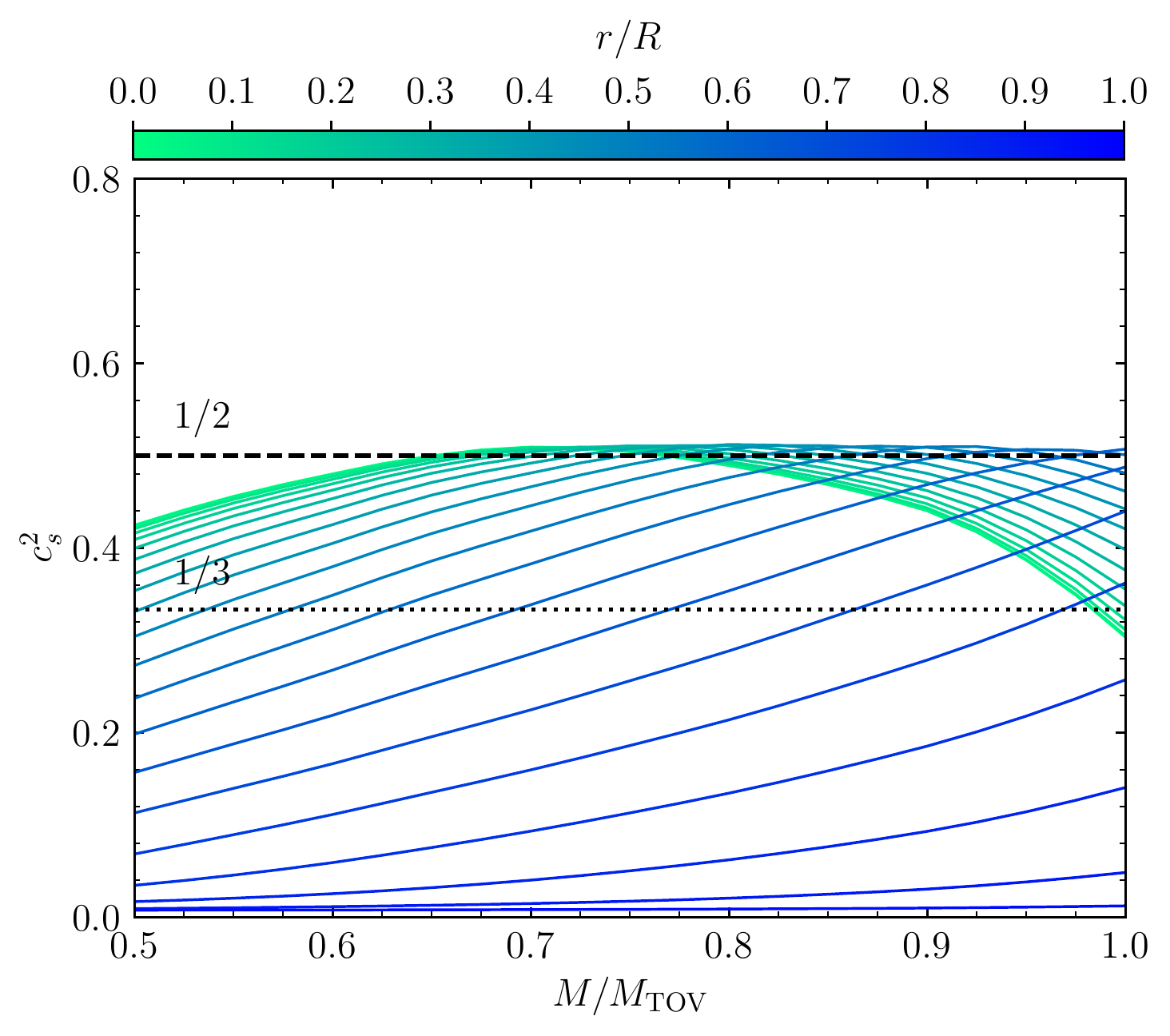}
  \caption{Mass dependence of the median sound speed for various
    different values of the normalized radial position location inside
    neutron stars as indicated by the colorbar. Black dashed and dotted
    lines mark half the speed of light and the conformal limit,
    respectively.}
\label{fig:cs2maxR}
\end{figure}

A few remarks on expression \eqref{eq:fitcs2} are in order. First, while
these functions are normalised in terms of the stellar radius and maximum
mass -- none of which are known at present -- they provide useful
information nevertheless. For instance, notwithstanding the limited
knowledge available now, it is already possible to conclude that a
neutron star with a mass close to the maximum mass of $\sim
2.2\,M_{\odot}$ \citep[see, \eg][]{Rezzolla2017} and an average radius of
$\sim 12\,{\rm km}$ \citep[see, \eg][]{Altiparmak:2022} will have a
maximum sound speed of $c_s \sim 0.7$ and at $\sim 8\,{\rm km}$ from the
center. This valuable information can already be included in
nuclear-theory calculations of new EOSs. Second, although a fit with
seven parameters may seem excessive, they are necessary to faithfully
represent the key features of the distribution, such as their values at
the center ($r=0$) and close to the surface of the stars ($r=R$), as well
as the location and value of the local maximum at intermediate
radii. Finally, the number of parameters could be reduced by imposing
analytic conditions on the approximate behavior of the PDFs in certain
limits (\eg using the vanishing slope at $r/R=0,1$). However, we have
preferred to be conservative and not to impose such constraints at the
cost of a larger number of parameters.

Figure~\ref{fig:cs2maxM} shows the behaviour of the median of the local
maximum sound speed $c^2_{s,{\rm max}}$ (red line) and of the sound speed
at the neutron-star center $c^2_{s,{\rm center}}$ (green line) as a
function of the (normalized) stellar mass. The red- and green-shaded
areas indicate the corresponding $95$\% credibility intervals, while the
black dashed and dotted lines mark the values $1/2$ and $1/3$,
respectively. In essence, Fig.~\ref{fig:cs2maxM} highlights that the
maximal sound speed in stars with $M/M_{\rm TOV}\lesssim0.7$ appears in
their center, because the red and green curves coincide for $M/M_{\rm
  TOV}\lesssim0.7$. On the other hand, for stars with $M/M_{\rm
  TOV}\gtrsim 0.7$ the two curves split: the sound speed at the stellar
center decreases monotonically towards the stellar interior reaching a
minimum of $c_{s, \rm center}^2\approx 0.3$ for the maximum-mass stars
($M/M_{\rm TOV}=1$). Conversely, the maximum sound remains at a constant
value $c_{s,\rm max}^2\approx 1/2$ for all masses $M/M_{\rm
  TOV}\lesssim0.7$. It is suggestive that this constant value is so close
to half of the speed of light, but it is difficult to invoke any
first-principle argument on why this should be the case.

Finally, in Fig.~\ref{fig:cs2maxR} we report the behaviour of the median
of the sound speed as a function of the (normalized) stellar mass at
different positions in the star, as indicated by the colorbar (these are
essentially cuts at different values of $r/R$ of the blue median surface
in Fig.~\ref{fig:cs2_3D}). In particular, the dark-green line represents
a cut of the median at the stellar center, while the dark-blue line shows
the behaviour at the stellar surface. In essence, Fig.~\ref{fig:cs2maxR}
highlights how the sound speed changes when going from the surface
(dark-blue line) towards the core (dark-green line) in stars as a
function of their mass. At $r/R\approx 0.65$ (light-blue line) the sound
speed reaches in the heaviest stars its maximum value $c^2_s \simeq 1/2$,
indicated by the black dashed line.

This shows that in the outer layers ($r/R\gtrsim 0.65$) of every star,
regardless of its mass, the sound speed is a monotonically increasing
function of the mass and has values $c_s^2<1/2$. On the other hand, the
sound speed in the core region ($r/R\approx 0$), shown in dark green, is
a non-monotonic function of the neutron-star mass. As a result, the
maximum sound speed at the stellar center, \ie $c_{s, \rm
  center}^2\approx 1/2$, is not attained in the lightest ($M/M_{\rm TOV}=
0.5$), nor in the heaviest stars ($M/M_{\rm TOV}= 1$), but rather at
intermediate mass $M/M_{\rm TOV}\approx 0.75$. Furthermore, the sound
speed at the center of light stars ($c_{s,\rm center}^2\approx 0.4$) is
actually larger than the corresponding value of the heaviest stars
($c_{s,\rm center}^2\approx 0.3$) that, as discussed above, experience a
significant softening.

\section{Summary and Conclusions}
\label{sec:conclusion}

We have studied the sound speed distribution inside neutron stars using a
large set of randomly generated EOSs that are consistent with nuclear
theory and perturbative QCD results in their respective ranges of validity and are in
agreement with astrophysical pulsar observations and gravitational wave
detections from binary neutron-star mergers. Our main result is a novel
and a scale-independent representation of the sound speed in a unit-cube
spanning the normalised radius $r/R$ and the mass normalized to the
maximum one, $M/M_{\rm TOV}$.

This innovative way of thinking about the sound speed has allowed us to
draw a number of general conclusions that increase our insight into the
quantitative and qualitative behavior in neutrons stars of the sound
speed in particular, and of dense nuclear matter more in general. More
specifically, we find that: 

\textit{(i)} close to the surface of the stars, the sound speed is small
and approximately independent of the mass. However, moving inside the
stars, the sound speed develops a non-trivial mass dependence, and for
$M/M_{\rm TOV}\gtrsim 0.7$, it changes from a monotonic to a
non-monotonic function of position $r/R$ with a single local maximum,
$c^2_{s, {\rm max}}$. The radial location of the maximum sound speed
depends on the mass and it is at the center of the star for light stars
but then moves to the outer layers of the star for heavy stars.

\textit{(ii)} the value of this local maximum becomes independent of the
mass for sufficiently large masses ($M/M_{\rm TOV} \gtrsim 0.7$) and
attains a constant value of $c_{s,\rm max}^2\approx 1/2$.

\textit{(iii)} the sound speed at the center of the stars, $c^2_{s,{\rm
    center}}$, also exhibits a non-monotonic behavior as a function of
the mass, with the maximum value of $c^2_{s,{\rm center}}$ being reached
by stars with intermediate mass, while the minimum value is not obtained
in the lightest stars but in the heaviest stars where it is $c^2_{s,{\rm
    center}} \simeq 0.3$, thus below the conformal limit. 

\textit{(iv)} using the sound speed as a measure for stiffness, we find
that light stars are soft in the outer layers ($r/R \simeq 0.5-0.7$) and
stiff in the core, while heavy stars have a soft core and stiff outer
layers. This is because the sound speed increases only slowly towards the
center in light stars, but rapidly in heavy stars where it approaches a
local minimum that is smaller than the conformal limit in the core.

\textit{(v)} finally, we provide a simple fitting formula for the median
sound speed and its confidence interval in the neutron-star
interior. This information can already be included in nuclear-theory
calculations of modern EOSs to constrain the behaviour of the sound speed
in those regions where nuclear-theory predictions have large
uncertainties.

In summary, the non-trivial behaviour of the sound speed as
  function of the radial position inside the stars can be seen as a probe
  to identify changes of the matter composition. In particular, the
  softening of heavy neutron-star cores points to the appearance of new
  degrees of freedom such as hyperons or deconfined quarks.

There are a number of possible extensions to our work. First, while our
method of constructing EOSs includes, in principle, cases that closely
resemble a first-order phase transition, they only represent a negligible
subset of our ensemble as we do not explicitly enforce them and hence
their statistical weight is rather small. It would be therefore
interesting to include models that, by construction, include a
first-order phase transition and study their impact on our
results. Second, in the current and also our previous
work~\citep{Altiparmak:2022}, we took the so-called frequentist approach
for the statistical interpretation of our results, that is, we impose the
constraints with a hard-cutoff neglecting their statistical nature. An
alternative approach to obtain a statistical interpretation would be a
Bayesian analysis where the statistics of the observational uncertainties
is taken into account. Finally, another interesting generalization of our
work is its extension to study how rotation -- and in particular rapid
rotation -- affects the properties of the sound speed in neutron
stars. We plan to address many of the these issues in ongoing and future
work.

\begin{acknowledgments}

We thank S.~Altiparmak for his contribution in the early stages of this
project and R.~Duque, E.~Fraga, T.~Gorda, J.~Margueron, C.~Moustakidis,
and S.~Reddy for useful discussions and comments. Partial funding comes
from the State of Hesse within the Research Cluster ELEMENTS (Project ID
500/10.006), by the ERC Advanced Grant ``JETSET: Launching, propagation
and emission of relativistic jets from binary mergers and across mass
scales'' (Grant No. 884631). C.~E. acknowledges support by the Deutsche
Forschungsgemeinschaft (DFG, German Research Foundation) through the
CRC-TR 211 ``Strong-interaction matter under extreme conditions''--
project number 315477589 -- TRR 211. The calculations were performed on
the local ITP Supercomputing Clusters Iboga and Calea.
\end{acknowledgments}

\appendix
\section{Imposing an upper bound on the maximum mass}
\label{sec:appendix_a} 
\setcounter{figure}{0}

\begin{figure*}[htb]
\center
\includegraphics[width=1\textwidth]{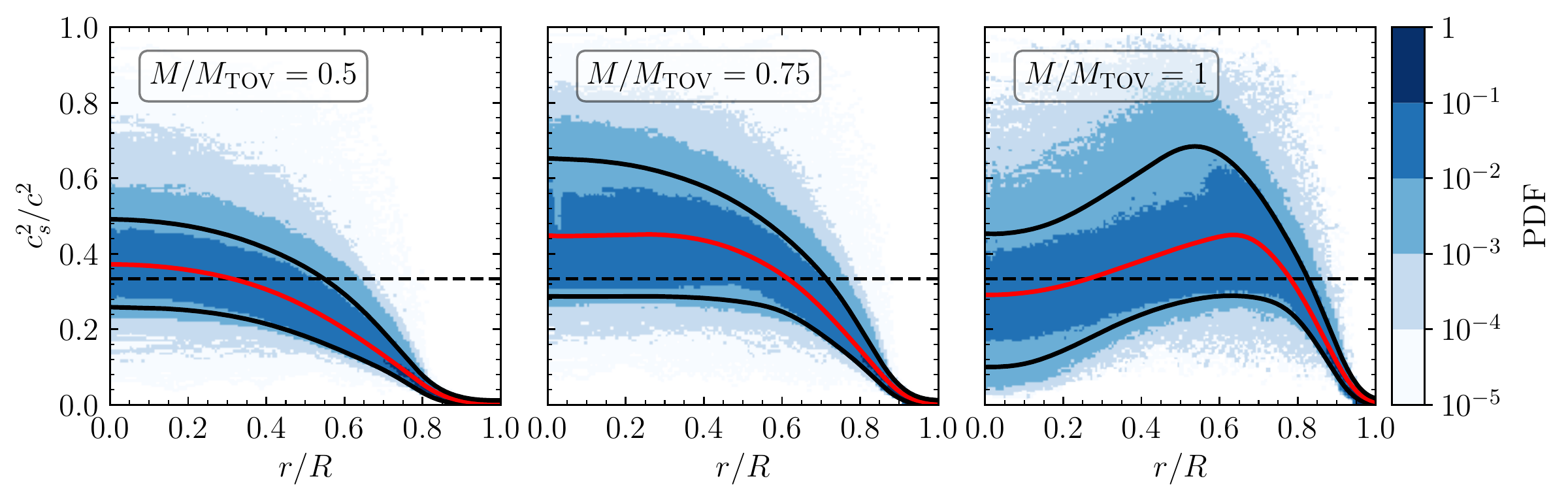}
\caption{Probability density functions for the sound speed as a function
  of the normalized radial coordinate $r/R$ for three fixed values of the
  masses $M=0.5\,M_{\rm TOV}$ (left), $M=0.75\,M_{\rm TOV}$ (middle) and
  $M=M_{\rm TOV}$ (right). Red lines represent the median of the
  distribution, black solid lines are the upper and lower bound of the
  $95$\% credibility interval, while black dashed lines indicates the
  conformal limit $c_s^2=1/3$. This figure is analogous to
  Fig.~\ref{fig:pdfMtov}, the only difference is that here an upper limit
  on the maximum mass $M_{\rm TOV}<2.16\,M_{\odot}$ is imposed.}
\label{fig:pdfBH}
\end{figure*}

In this Appendix we analyze the impact on the sound-speed PDF of an
additional constraint obtained by imposing and upper bound on the maximum
neutron-star mass, namely, $M_{\rm TOV}\lesssim
2.16^{+0.17}_{-0.15}\,M_{\odot}$. as proposed by~\citet{Rezzolla2017}
\citep[see also][for a discussion on the constraints derived from
  GW190814]{Nathanail2021}. Such a constraint is obtained using the
detection of the gravitational-wave event GW170817 and the modeling of
the kilonova signal of GRB170817A, together with quasi-universal
relations between the maximum masses of uniformly rotating and
non-rotating stars~\citep{Breu2016}. Figure~\ref{fig:pdfBH} shows the
sound-speed PDF in a way that is similar to the three top panels of
Fig.~\ref{fig:pdfMtov}, with the difference that the PDF here is obtained
after imposing an upper limit on the TOV-mass $M_{\rm
  TOV}<2.16\,M_{\odot}$.
A rapid comparison between Figs.~\ref{fig:pdfBH} and \ref{fig:pdfMtov}
reveals that the overall features of the sound speed remain
unchanged. The most significant difference is that the distributions are
systematically narrower, simply because the stiff EOSs with large sound
speeds are now penalized and rejected. This is particularly clear in the
panel for heavy stars (right), where the sound speed increases less
rapidly in the outer layers of the star. Interestingly, the value of the
local maximum sound speed, $c_{s,\rm max}$, varies only minimally,
changing from $c_{s,\rm max}\simeq1/2$ in the absence of a maximum-mass
constraint, to $c_{s,\rm max}\simeq0.45$ when a maximum-mass constraint
is imposed.

\section{Comparison to microphysical models}
\label{sec:appendix_b} 

\begin{figure}[htb]
\center
\includegraphics[width=0.48\textwidth]{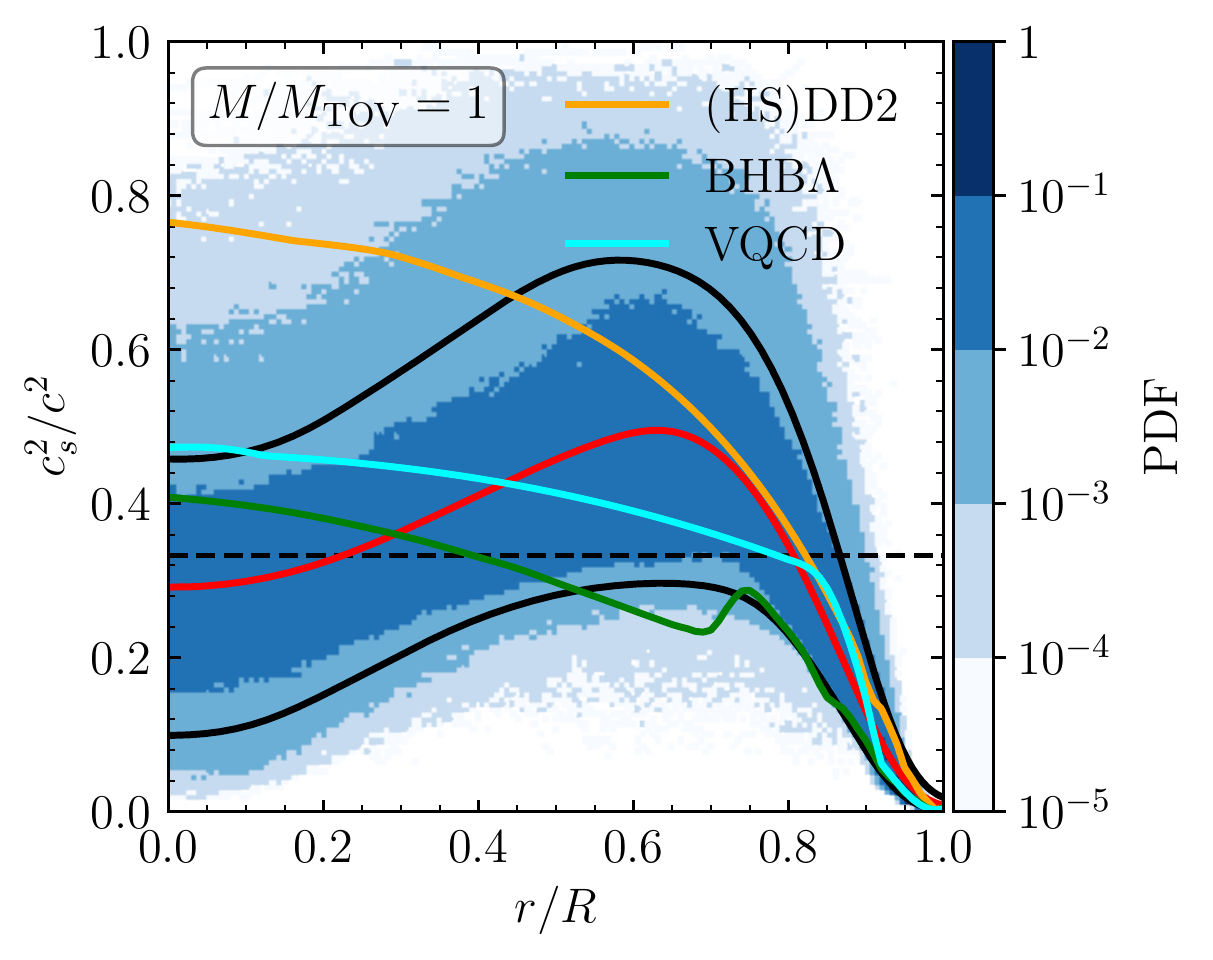}
\caption{Probability density function of the sound speed for $M=M_{\rm
    TOV}$. The red line represents the median of the distribution, black
  solid lines are the upper and lower bound of the $95$\% credibility
  interval, while the black dashed line indicates the conformal limit
  $c_s^2=1/3$. Orange, green and cyan lines report the cold beta
  equilibrium sound speed for the HS(DD2)~\citep{Hempel:2009mc,
    Typel:2009sy}, the BHB$\Lambda$~\citep{Banik2014} and the
  intermediately stiff VQCD~\citep{Demircik:2021zll} EOS, respectively.}
\label{fig:compare}
\end{figure}

In this Appendix we compare our sound-speed PDF with three microphysical
models that have been widely used in the literature to study the
properties of neutron stars and their mergers. Figure~\ref{fig:compare},
in particular, compares the radial behaviour of the sound-speed PDF in
maximally massive stars with the corresponding sound speeds obtained with
these EOSs.

The yellow line shows the sound speed of a pure nuclear-matter model, the
Hempel-Schaffner-Bielich (HS) EOS with DD2 relativistic mean field
interactions~\citep{Typel:2009sy,Hempel:2009mc}. This EOS is known to be
relatively stiff as can be seen from the large sound speed in the
neutron-star interior. Interestingly, for $r/R\gtrsim 0.7$ the sound
speed from this EOS agrees very well with the median sound speed (red
line) from our approach, even though our construction does not use any
input from the (HS)DD2 EOS.  However, deeper inside the star, our result
is very different from (HS)DD2 and gives less than half the sound speed
in the neutron-star core. We should note that the (HS)DD2 EOS gives a
relatively high maximum mass of $M_{\rm TOV}\approx 2.5\,M_{\odot}$ and a
tidal deformability of $\Lambda_{1.4}\approx690$ for
$1.4\,M_{\odot}$-stars, which is in tension with the upper bound
$\Lambda_{1.4}\lesssim 580$ derived from the inspiral part (low-spin
prior) of GW170817~\citep{Abbott2018a}. This means (HS)DD2 only provides
a good description for dense matter $n\lesssim n_s$ in light stars, but
is probably too stiff to describe the properties of heavy neutron stars
with cores several times denser than $n_s$.

The green line, on the other hand, corresponds to the
Banik-Hempel-Bandyopadhyay (BHB$\Lambda$) EOS~\citep{Banik2014}, which is
an extension of the (HS)DD2 EOS with $\Lambda$-hyperons, i.e., particles
that contain strange quarks. One of the characteristic features of
hyperonic degrees of freedom is that they lead to a softening of the EOS,
which can be seen from the local minimum in the sound speed around
$r/R\approx 0.8$. The maximum mass of this EOS is relatively low $M_{\rm
  TOV}\approx2.0\,M_{\odot}$ and therefore in tension with the mass
measurement $M\gtrsim 2\,M_{\odot}$ of heavy pulsars. Hence, the EOS is
probably too soft to account for the properties of light stars and the
outer layers of heavy stars as can be seen from the large difference
between the red and green curve at $r/R\gtrsim 0.5$ in
Figure~\ref{fig:compare}.

Finally, the cyan line corresponds to the intermediately stiff V-QCD EOS
of \citet{Demircik:2021zll}, that combines the traditional nuclear theory
(HS)DD2 and the Akmal-Pandharipande-Ravenhall (APR)
EOS~\citep{Akmal:1998cf} EOSs at low densities with a string theory
inspired model for QCD to describe dense baryonic and quark matter. This
model gives $M_{\rm TOV}=2.14\,M_{\odot}$ and $\Lambda_{1.4}=511$ and
therefore satisfies conveniently the two-solar mass and tidal
deformability constraints. Furthermore, it has been verified
recently~\citep{Tootle:2022pvd} via binary neutron-star merger
simulations, that this EOS is not excluded by the expected one
second-long lifetime~\citep{Gill2019} of the post-merger remnant of
GW170817. However, although this model passes all currently known
theoretical and observational constraints, it is also not able to account
for the softening of the core predicted by the red curve.

In summary, none of the three very different microphysical EOSs shown in
Fig.~\ref{fig:compare} is able to describe the behavior of the sound
speed predicted by our model-agnostic sampling approach. The explanation
is simple and reflects the difficulties that nuclear-theory calculations
have in matching the high-density, perturbative QCD constraints.
It is remarkable that EOSs such as those considered in
  Fig.~\ref{fig:compare} -- that are in good agreement with all currently
  known observational constraints -- have radial profiles of the sound
  speed that are quite different those predicted from our agnostic
  approach. This fact shows that the traditional methods of constraining
  EOS models only with global neutron-star properties such as the maximum
  mass, the radii and tidal deformabilities, might not be
  sufficient. Rather, additional information on the properties inside the
  stars is needed and can provide further non-trivial constraints. Hence,
  the importance of our Eq.~\eqref{eq:fitcs2} is to provide a novel
  constraint in terms of the radial distribution of the sound speed
  inside stars of a given mass. Using this constraint will ensure that
  the newly suggested EOSs will not only satisfy the astrophysical
  constraints, but are also compatible with the perturbative QCD
  constraints at much larger densities. We are not aware of a comparable
  constraint in the literature.

\section{Comprehensive description of the fitting}
\label{sec:appendix_c} 

In Table~\ref{tab:fit} we provide the numerical values of the fitting
coefficients for the results shown in Fig.~\ref{fig:pdfMtov};
these coefficients reflect the astrophysical constraint imposed
  here of $M_{\rm TOV}\gtrsim2~M_{\odot}$ and their values may change
  slightly if this constrain is increased. As briefly mentioned in the
main text, some of the parameters in Eq.~\eqref{eq:fitcs2} can in
principle be related by strictly imposing the approximate behavior of the
distribution in certain limits.  For example, the numeric results have
approximately vanishing slope at $r/R=0,1$ and the value of the local
maximum of the median at $0<r/R<1$ turns out to be very close to the
value $1/2$ in stars with $M/M_{\rm TOV}>0.7$. Imposing these constraints
would allow to express three of the coefficients in terms of the
remaining four, but at the same time make the formal expression
\eqref{eq:fitcs2} more involved. In addition, the upper and lower bounds
of the credibility interval do not show the saturation behavior of the
local maximum seen in the median such that it would not be possible
anymore to expressed them by the same fitting law. We also remark that
the sound speed loses the local maximum for $M/M_{\rm TOV}\lesssim0.7$
and becomes monotonic, allowing for a simpler fitting law with fewer
parameters. This can be directly seen from the values of $\alpha$
provided in Table~\ref{tab:fit}, which become small compared to the other
parameters, meaning that the first exponential function in
Eq.~\eqref{eq:fitcs2} can be neglected, reducing the number of relevant
parameters to five.

\tabcolsep=0.1cm
\begin{deluxetable}{ccccccccc}[h!tb]
\tablecolumns{9}\label{tab:fit}
\centering
\tablecaption{Numeric values of the fitting parameters in
  Eq.~\eqref{eq:fitcs2} for the median sound speed and the lower and upper
  bound of its $95$\% confidence interval for three different values of
  $M/M_{\rm TOV}$ used in Fig.~\ref{fig:pdfMtov}.}
\tablehead{\colhead{$\tfrac{M}{M_{\rm TOV}}$} & \colhead{value} &
  $\alpha$ & $\beta$ & $\gamma$ & $\delta$ & $\epsilon$ & $\zeta$ & $\eta
  $}
\startdata
      & lower  & $0.02$ & $0.14$ & $0.47$ & $-12.00$ & $-2.60$ & $15.00$ & $12.00$\\
  0.5 & median & $0.01$ & $0.21$ & $0.42$ & $-18.00$ & $-1.60$ & $7.70$ & $6.00$\\
      & upper  & $0.01$ & $0.31$ & $0.37$ & $-27.00$ & $-1.10$ & $6.60$ & $5.00$\\
  \hline
      & lower  & $0.09$ & $0.15$ & $0.58$ & $-9.80$ & $-3.40$ & $13.00$ & $11.00$\\
  0.75& median & $0.08$ & $0.25$ & $0.55$ & $-10.00$ & $-1.80$ & $9.80$ & $8.40$\\
      & upper  & $0.09$ & $0.15$ & $0.58$ & $-9.80$ & $-3.40$ & $13.00$ & $11.00$\\
  \hline
      & lower  & $0.10$ & $0.03$ & $0.54$ & $-6.40$ & $1.50$ & $14.00$ & $12.00$\\
  1   & median & $0.18$ & $0.14$ & $0.71$ & $-6.30$ & $-1.50$ & $10.00$ & $8.90$\\
      & upper  & $0.19$ & $0.23$ & $0.65$ & $-7.80$ & $-0.77$ & $12.00$ & $11.00$\\
  \enddata
\end{deluxetable}

\bibliographystyle{aasjournal}
\bibliography{aeireferences}

\end{document}